# QUEST FOR A NUCLEAR GEOREACTOR[1]

R.J. de Meijer, E.R. van der Graaf and K.P. Jungmann
Kernfysisch Versneller Instituut, Rijksuniversiteit Groningen
Zernikelaan 25, 9747AA Groningen, the Netherlands
(12 augustus 2003)

**Abstract**
Knowledge about the interior of our planet is mainly based on the interpretation of seismic data from earthquakes and nuclear explosions, and of composition of meteorites. Additional observations have led to a wide range of hypotheses on the heat flow from the interior to the crust, the abundance of certain noble gases in gasses vented from volcanoes and the possibility of a nuclear georeactor at the centre of the Earth. This paper focuses on a proposal for an underground laboratory to further develop antineutrinos as a tool to map the distribution of radiogenic heat sources, such as the natural radionuclides and the hypothetical nuclear georeactor.

**Introduction**
In a time when astronauts orbit the Earth and visit the Moon, and mankind has brought vehicles to Mars and telescopes into orbit, we seldom realise that we have penetrated the Earth by only 10km, a distance smaller than we commute daily to our work or equivalent to the cruising altitude of airplanes. Consequently we know, except from seismic information and study of the composition of meteorites, little about the interior of our planet.

The current understanding of the interior of our planet starts from seismological investigations by e.g. Oldham (1906) and Gutenberg (1914) that lead to the hypothesis that up to half the radius of the Earth is occupied by a fluid core. From the interpretations of earthquakes Lehmann (1936) recognised a small, solid inner core. The fact, that meteorites consist of nickelferous iron, lead to the assumption that the fluid core of the Earth consists of molten nickel-iron. Density information stems from the work of Birch (Birch, 1952). He hypothesised from seismological models and knowledge on high-pressure equations of state that the outer core was composed of a liquid iron alloy and an inner solid core of crystalline iron. The melting temperature of the alloy at the respective pressure defines the boundary. Estimates for this temperature at the pressure of 330 GPa range between 5000 and 6000 K. Figure 1 presents the principal divisions and physical states of the Earth's interior. The absence of shear velocity $V_s$ of earthquake waves is the basis for a fluid core. The density curve shows in addition to the major changes at the principal sections steps in the upper mantle at about 420km and 660km depth.

Heat loss from the core depends on the radial temperature gradient at the boundary of core and overlying mantle and is strongly related to mantle dynamics. Here exists a large uncertainty in $\Delta T$, due to the temperature at the inner-core boundary: $\Delta T$= 1000 to 1800 K (Anderson, 2002) over a layer of a few hundred kilometres (Lay *et al.*, 1998). Including the thermal conductivity of the mantle silicates yield a heat flow of 0.04-0.08 W/m$^2$, leading to a total heat flow from the core of 6-12TW (Buffett, 2003); a considerable part of the estimated total heat flow from the Earth of 40-50TW. The total heat flow at the core-mantle boundary raises vital questions on the thermal evolution of the core and its heat sources in relation to power required to maintain the magnetic field. Radiogenic elements like $^{40}$K are thought to play an essential role (Rama Murthy *et al*, 2003).

---

In his paper Buffett concludes that "the thermal state of the core remains unclear and that better knowledge of the partitioning of all radiogenic elements between various reservoirs in the planet will help to reduce some ambiguity".

Another ambiguity exists on the chemical composition of the various compartments or reservoirs and especially the core. In general one assumes that there is a liquid Fe-Ni alloy core, surrounded by a lower and upper mantle and covered by a crust. The bulk composition of the Earth is usually assumed to be the same as that of chondritic meteorites. Within this assumption subsequent hypotheses are made to account for observations at the Earth's surface. An intriguing issue is the presence of Helium in our atmosphere and in particular its isotope $^3$He. Whereas $^4$He is continuously produced by alpha decay, the only way to obtain $^3$He is either as a primordial relict (e.g. Seta *et al.*, 2001) or by decay of tritium. For the primordial relict the assumption has to be made that the mantle contains a degassed and a, deeper lying, less-degassed reservoir. The former one shows up at the mid-ocean ridge basalts, the latter one in mantle-plumes basalts. Mantle plumes with extreme high $^3$He/$^4$He ratios are found at some oceanic islands such as Iceland, Hawaii, Samoa and Galapagos (Kurtz and Geist, 1999) One assumption commonly made in interpreting noble gas data from mantle plumes is that the source of mantle plumes is relatively non-degassed lower mantle material. Under this assumption, high $^3$He/$^4$He ratios indicate plume-like upwelling, since the deep Earth is believed to be a source of primordial $^3$He with a relatively low time-integrated (U+Th)/He ratio (Georgen *et al.*, 2003) In oceanic islands not only high $^3$He/$^4$He ratios are found but also normal mid-oceanic island values. This is explained by assuming mixed reservoirs (Stuart *et al.,* 2003).

Recently Bercovici and Karato (2003) proposed a filtering of the mantle at the 410km discontinuity of the density (see Figure 1). They propose that the ascending mantle rises out of a transition zone, between the 410 and 600km discontinuities, into the upper mantle above 410km. The material undergoes dehydration-induced partial melting which filters out incompatible elements, including He and other noble gases. They propose that this filter model can explain geochemical observations without the need for isolated mantle reservoirs. This model could bridge the gap between geochemists supporting a two-layer model at a boundary of 660km and seismologists, supporting a whole-mantle model of circulation (Hofmann, 2003).

Recently a possible explanation for some of these questions was given by hypothesising a 8km diameter, nuclear georeactor at the centre of the Earth. The hypothesis for such a reactor originates from the work of Herndon (1992) in applying Fermi's nuclear reactor theory to demonstrate the feasibility of planetary scale nuclear fission reactors. Such reactors could be the energy source of the giant outer planets, three of which radiate about twice the energy they receive from the Sun. Subsequently he extended the feasibility of such a reactor at the centre of the Earth as a contributive source to geodynamic processes like plate movements (Herndon, 1993, 1994, 1996). Such a reactor started in the same way as the natural reactors at Gabon, but is so large that it breeds its own $^{235}$U and should have a power production of 3 to 10TW. In such a reactor tritium is produced via ternary fission. Calculations at Oak Ridge National Laboratory (Hollenbach and Herndon, 2001) show that a planetary-scale nuclear reactor can operate over the lifetime of the Earth as a breeder reactor and can produce substantial tritium (decaying to $^3$He) to explain the high $^3$He/$^4$He ratios observed in oceanic basalts and fumes of volcanoes at Iceland and Hawaii. Seifritz (2003) shows that the operation of such a breeder reactor is consistent with our knowledge on breeder reactors and corresponds to a stable state.

The possibility of a nuclear georeactor is linked to the state of oxidation in the deep interior of the Earth. Herndon has convincing arguments for a state of oxidation like an enstatite chondrite, different from the more oxidised, ordinary chondrites considered by Birch.

As a consequence of the highly reduced state some so-called lithophile elements including some Si, Mg, Ca, U and possibly Th occur in part of the core. These elements, tending to be incompatible in an iron alloy, are expected to precipitate at relative high temperatures. Due to their density MgS and CaS will float to the core-mantle boundary, whereas uranium sulphide (US) and nickel silicide will sink to the Earth's centre.

At pressures that prevail in the core, U and Th, being high-temperature precipitates and the densest substances would tend to concentrate in the Earth core by the action of gravity. In that process it will ultimately form a fissionable, critical mass. Fission produces less (half) dense fission products that tend to separate from the more dense actinides. In this way a critical reactor condition can maintain.

According to Herdon (1993) and Hollenbach and Herndon (2001) the frequent, but irregular variability in intensity and direction of the Earth's magnetic field may be understandable from such a fission reactor. The production of fission products counteracts the operation of the reactor and if the rate of production exceeds the rate of removal by gravitational diffusion, the output of the reactor will decrease and may even shut down, leading to a diminishing and ultimately disappearing of the Earth's magnetic field. As fission products diffuse out of the reactor region and actinides diffuse inwards, the reactor restarts and the geomagnetic field re-establishes itself, either in the same or in the reverse direction. The coupling between the georeactor and the geomagnetic field cannot be directly (Hoyng, 2003) and has to proceed through changing heat-flow patterns in the core and ultimately in the mantle.

Although the georeactor hypothesis seems to be able to explain, in principle, phenomena such as elevated $^3$He/$^4$He ratios and reversal of the geomagnetic field also some questions remain about the specific mechanisms involved. Regarding the working of the georeactor it is assumed that fission products are separated from the fuel by diffusion or by buoyancy effects. For both processes it still has to be shown if they are effective enough. First estimates for diffusion, based on an extrapolation (using Arrhenius law) to core temperatures of diffusion coefficients and activation energies for helium in apatite (Dunai, 2000) indicate that transport over a distance of 1 km in a solid metal inner core will take 1 Ma. This is probably at least an order of magnitude to slow to explain geomagnetic reversals every 200 000 years by fission products drifting outwards and so cleaning up the core for a reactor restart. To estimate transport velocities due to buoyancy detailed calculations are needed but these velocities are expected to be insignificant because of the micro-gravity conditions in the inner part of the core (Seifritz, 2003).

Also it is difficult to imagine a sufficiently large outward flux of $^3$He (needed to explain elevated $^3$He/$^4$He ratios) produced inside the georeactor because an intact solid inner core will form an almost impenetrable barrier for transport. Moreover also the heat produced by the georeactor has to be removed through the solid inner core and normal heat conductivities for solids are not large enough to prevent the reactor from heating up to very high temperatures.

It should be taken into consideration that all these estimates are based on extrapolating transport coefficients at ambient conditions to the high temperatures and pressures inside the core. Moreover these estimates assume a uniform metallic core being one solid piece. A more granular structure of the inner core with a structure of "pores" or fissures will allow a larger and faster transport of gaseous substances and heat. In such a case it is hard to imagine that the core will be uniform and likely preferential pathways may exist, which act as chimneys and may cause a non uniform heating of the inner-outer core boundary.

**Antineutrino's as a tool to probe the Earth's interior**

Because of the work of Herndon it is no longer a question what the distribution is of U and Th in the core and the mantle but it is also required to find out whether they decay along their decay series or by fission. One of the few methods to investigate the distribution of natural radionuclides in various reservoirs of the Earth and/or the existence of a nuclear georeactor are antineutrino's produced in β-decay and/or fission, respectively. Fortunately the decay and fission can be distinguished by the energy of the antineutrino's; 2-3 MeV for decay and up to 10 MeV for fission. To map the U-Th distribution and to localise a nuclear georeactor directional information is required in antineutrino detection.

The science and technology of detecting antineutrino's, $\overline{v}_e$, is well established. Like in many experiments we propose to use the detection reaction based on inverse β-decay: $\overline{v}_e$ + p → $e^+$ + n. The visible energy of the positron signal directly provides the $\overline{v}_e$ energy (E(MeV)= E($\overline{v}_e$)-1.8+1.022= E($\overline{v}_e$)-0.78. The signal can be tagged by the signal produced after several tens of microseconds by the thermalised neutron captured by hydrogen in the aromatic organic liquid scintillator. The delayed coincidences suppress background enormously and the chance coincidence rate in a kiloton scintillator mass detector such as installed at Kamioka, Japan, can be limited to several events/year (Rhagavan, 2003). This corresponds to a sensitivity limit of an antineutrino flux $\Phi(\overline{v}_e)_{min}$~$10^4$ $cm^{-2}$ $s^{-1}$.

In β-decay of members of the decay series of U and Th antineutrino's, $\overline{v}_e$, are produced. The antineutrinos from the geo-reactor will be identical in energy spectrum as those produced in power reactors. These neutrinos can reach the surface of the Earth essentially without any interaction. One of practical methods to detect antineutrinos is by the inverse β-decay: $\overline{v}_e$ + p → $e^+$ + n with a threshold of 1.8 MeV. In the U series the β-decay of $^{234m}$Pa, $^{214}$Bi with Q-values of 2.29 and 3.26MeV, respectively and $^{228}$Ac, $^{212}$Bi and $^{208}$Tl with Q= 2,11, 2,25 and 1.8 MeV, respectively, can contribute. The detection of the antineutrinos is favoured by the Coulomb fields of these high-Z nuclei, which enhances the energy distribution towards the high-energy side of their spectrum (Raghavan *et al.*, 1998). The specific emission probabilities above threshold becoming 0.40 and 1.6 per decay for U and Th, respectively. The intensity of the U/Th antineutrinos depends on the distribution of these natural radionuclides relative to the position of observation. In the crust the concentrations of these radionuclides are more than an order of magnitude higher than in the upper mantle. The thickness of the crust varies from about 5km at the ocean floors to about 50km under the Himalayas. The mantle is much thicker but the distribution of the radionuclides is unknown. Some ideas about the distribution depend on the assumed model. The strength of the signal at various sites on Earth will therefore be an indication of the distribution of these radionuclides in the compartments of the Earth, even if no directional information can be derived from the antineutrino detection. According to calculations by Raghavan *et al.*,1998 40% of the signal will come from sources in the crust within about 450km, 70% from within 1200km and 90% from about 6000km.

For reactors, either power reactors or a geo-reactor, the antineutrino spectrum follows from their mean fuel composition, the numbers of antineutrinos per fission event and their spectrum ( see Achkar *et al.*, 1996 and references therein). Also for these neutrinos the flux depends on the distance between power reactor and detector. The existing underground laboratories at Gran Sasso, Italy, and Kamioka, Japan, are both situated on the crust and, especially Kamioka, near power stations.

A 3-10TW georeactor would yield at any point near the surface a flux $\Phi(\overline{v}_e)_{geo}$~1-3·$10^5$ $cm^{-2}$ $s^{-1}$ and is an order of magnitude larger than the estimated detector background of

~$10^4$ cm$^{-2}$ s$^{-1}$. So the detection of antineutrino's from the core is a valid proposition provided the background is sufficiently low. The background arises primarily from operating commercial power reactors within several 1000km and U, Th and K in the crust of the Earth, mainly located at the continents. Nominally a georeactor is spectrally nearly indistinguishable from power reactors, but the georeactor provides a strongly directional signal. Geo U/Th signals cut off at about $E_{e+}$ of about 2.5MeV. Model calculations indicate that such measurements are feasible in an underground laboratory, provided the background due to the geological formation and antineutrinos produced in nuclear power reactors is sufficiently low. This condition excludes observation in the existing underground laboratories such as Borexino at Gran Sasso, Italy, and Kamland at Kamioka, Japan and favours locations such as Hawaii, the Aleuts and the Antilles. Figure 2 shows the positron energy spectrum for Hawaii and Kamioka (taken from Raghavan, 2002). It shows that the background at Kamioka is too large for proper detection. This is recently confirmed by measurements with the Kamland detector yielding 9±6 geo-antineutrino's from an exposure of $1.4 \cdot 10^{31}$ protons year (Fiorentini *et al.*, 2003).

In a low-background environment one would expect from a ten times larger fiducial volume a signal of some one hundred events per year in a detector following the basic design of the Kamland or Borexino experiments (see figure 4). Directional information could be extracted from the recoiling neutron in the reaction $\overline{\nu}_e + p \rightarrow e^+ + n$, for which the recoil angle on average is $\langle \Phi_{recoil} \rangle = \arccos(2/(3A))$ with A the mass number of the scattering material. For the antineutrinos of geophysical interest the neutron travels of order few cm between the locations of positron creation and respectively neutron absorption. An antineutrino detector with such position resolution could follow the principle of the ANTARES experiment (see figure 5). This should add significant information about the location of the main antineutrino source.

**Proposed underground laboratory at Curaçao.**

We like to propose an underground laboratory to investigate the internal state of the Earth. We propose such a laboratory to be built on e.g. Curaçao. The geology of Curaçao, as studied by Klaver (1987) indicates that Curaçao is a mantle plume originating from the boundary of the core and the mantle, some 80 Ma ago. Sample analysis by Klaver (1987) indicate that indeed the Curaçao basalt is more than an order of magnitude lower in K,Th and U compared to sands in e.g. The Netherlands. Parts of the island of Curaçao, represent the magma after 5km of material has been eroded.

The proposal foresees in a step by step approach, starting from re-examination of surface rock formations, drilling for deeper material and finally excavating an underground laboratory. Curaçao is more than 1000km away from the Florida power reactors and from the mountain ranges of the Andes (see Figure 3). It provides therefore not only a very low background of natural radionuclides, but since it is surrounded by a considerable mass of ocean water, the antineutrino flux from crustal sources and power reactors will be strongly reduced. That means that the laboratory will be especially sensitive to antineutrino sources from the mantle and a possible geo-reactor. We expect that the calculation for Hawaii will be more or less indicative for Curaçao.

In addition to the antineutrino detection, such a very low background laboratory can be a base for low-background experiments such as the search for double β-decay. The creation of such an underground laboratory will be quite unique and will also include some technological challenges. One of them is drilling into basalt; the other is the development of low-energy dissipating electronics. The first is to the direct interest of energy generating and distributing companies that search for storage of gas in underground caverns (Tractebel, 2000). Despite Curaçao being a plume sticking out of the ocean floor and cooled by ocean water at a depth of about 1km, high temperatures are expected. In the underground laboratory we expect a high density of electronic devices in the detector setup. Since additional cooling is complicated, low-energy dissipating electronics will be required. Low-energy dissipating electronics is directly linked to the telecommunication technology and its industry.

It should be noted that the geological formation of the island of Curaçao would allow calibration of an antineutrino detector with the aid of nuclear power driven vessels which could be positioned at various locations and at variable distances from the detector.

Building and operating such an underground laboratory requires education and training of local people and can contribute to alleviate some of the economic and social problems on the Antilles. A project like indicated here, not only exceeds the capacity of a single institution because of its inter- and intradisciplinary character. Various branches of Physics (geophysics, plasma physics, nuclear physics and high-energy physics) form a basis for an even broader Science platform including several branches of geo-sciences and technology. Its goal, size and financial aspects also exceeds the national level and the project should preferably be executed on a European or even larger scale. This makes the financing of such a project feasible and could be the start of e.g. a European Earth Agency. The advantage of this proposal is that it starts relatively small and will expand as Science and finance interests grow.

As initial steps the re-examination of the surface rock formation and a deep drill are foreseen. This includes the location of the centre of the magma plume. Already in each of the stages valuable information is expected from the analysis of the cores, such as $^3$He/$^4$He and $^{10}$Be/$^9$Be ratios as well as ratios of stable isotopes and other radionuclides like $^{26}$Al. These results will already help to test the physics and geochemistry aspects of the various models. In the meantime the design of the antineutrino detector, starting from the Kamland detector, but tuned to the detection of antineutrinos from the natural radioactivity and the hypothetical geo-reactor will take place. This approach also illustrates why the KVI can act as the catalysing institute: it has the expertise of low-energy nuclear physics for the antineutrino detection and the Nuclear Geophysics Division has an expertise in analysing bore cores on low level natural radioactivity. Moreover the KVI has a vast network connecting to (nuclear) physics and geo-science institutes in the Netherlands and abroad.

By this paper we invite interested people with expertise in one of the fields, who like to participate in this proposed project to contact us.

Finally: we like to refer to this project as CURACAO (Curacao Underground Research Arena for Core Antineutrino Observations).

The authors like to acknowledge the stimulating discussions with J.M. Herndon, P. Hoyng, G. Th. Klaver and A.E.L. Dieperink.

**Figure Captions**

Figure 1. Schematic representation of the principal partitions and physical states of the Earth's interior. The compressional and shear velocities of earthquake waves, presented in the right panel, are indicated by $V_p$ and $V_s$, respectively. In the right panel the density as function of depth is presented. (Figure taken from Herndon, 1980).

Figure 2. Calculated positron energy spectrum for two locations: Hawaii and Kamioka. The two locations differ quite strongly in the flux from U/Th antineutrinos and from the antineutrinos produced in nuclear power stations. The situation of Hawaii will be quite similar to the one expected at Curaçao. (Figure taken from Raghavan, 2002.)

Figure 3. World map with nuclear power stations. The stations at Cuba and Puerto Rico, are not operational. (Figure taken from the International Nuclear Safety Center at Argonne National Laboratory http://www.insc.anl.gov).

Figure 4. Schematic presentation of the Kamland detector at Kamioka, Japan. The central part consists of a liquid scintillator sphere of several metres diameter, surrounded by a sphere of photo-multipliers. The outer detector contains the Cherenkov counters as active shielding against muons. (Figure taken from
http://kamland.lbl.gov/FiguresPlots/kamlandfigs_paper2002/xdetector5-thumb.png).

Figure 5. Artistic view of the detectors of the ANTARES experiment. Strings of phototubes are immersed in the active detector volume. This concept allows also sufficient position resolution for antineutrinos in a large scintillator, underground detector by determining the recoil direction of neutrons in the inverse β-decay process.
(Figure taken from http://antares.in2p3.fr/Gallery/index.html).

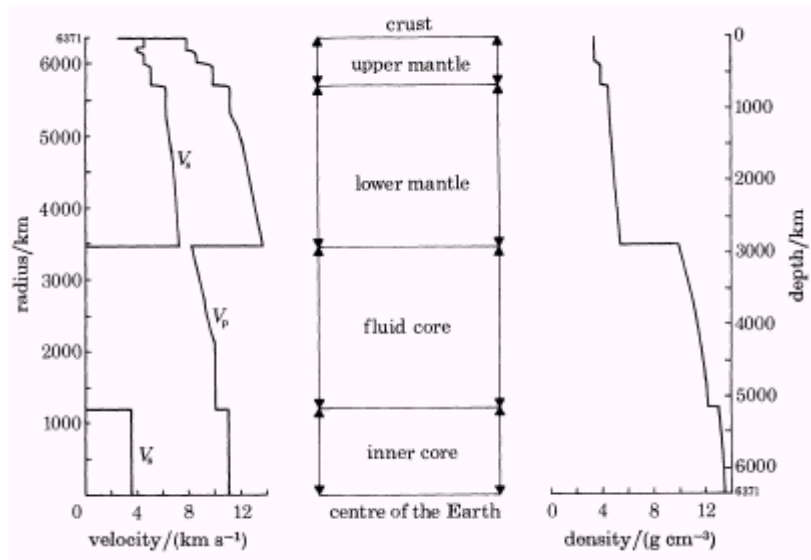

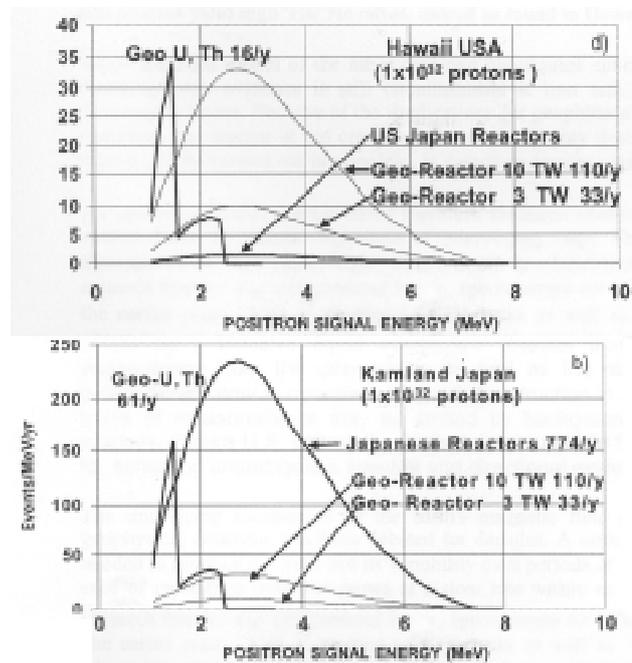

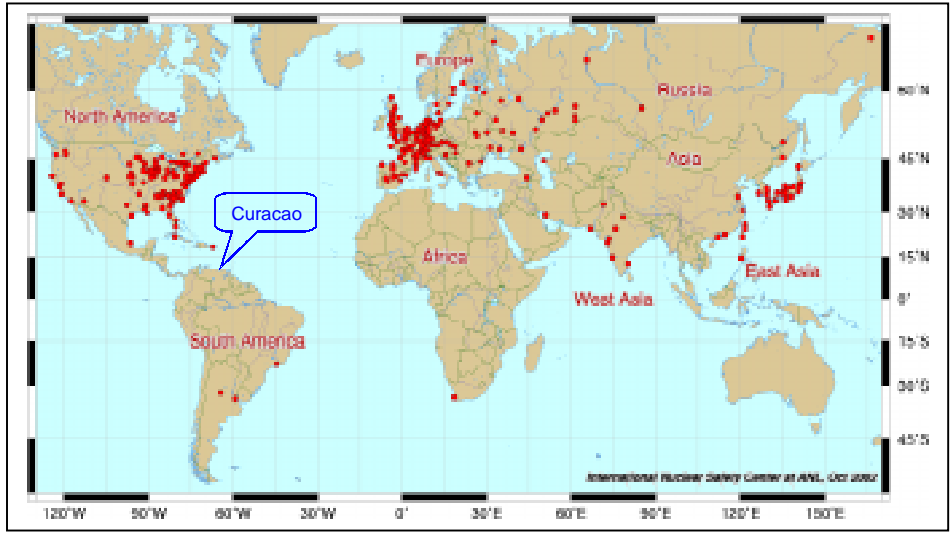

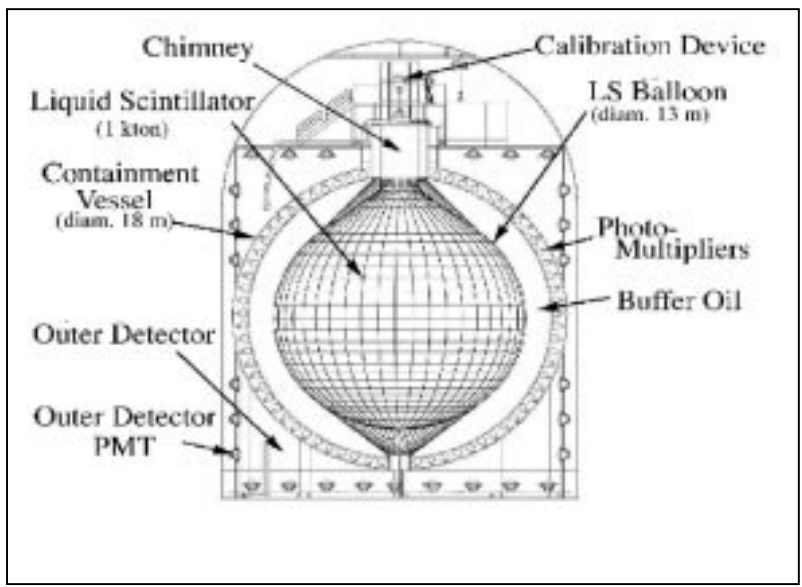

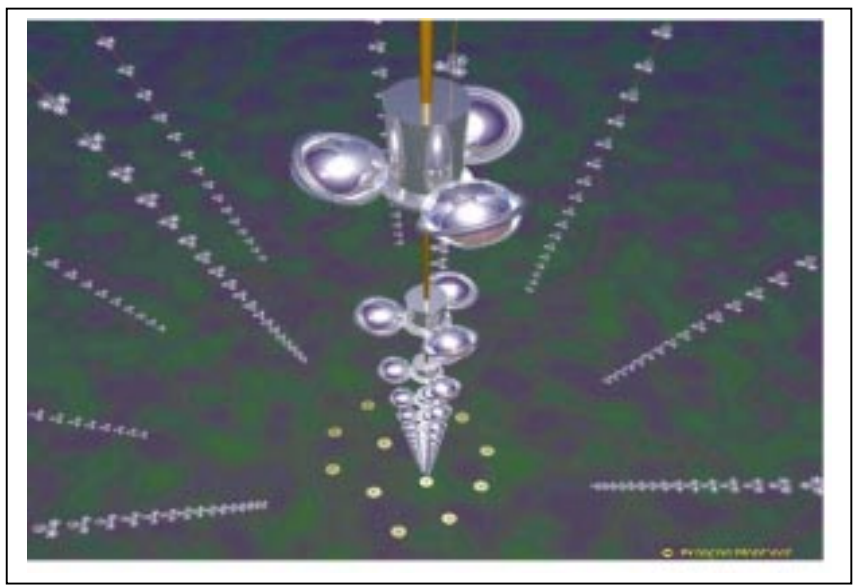